\documentstyle[preprint,aps,12pt]{revtex} 
\draft
\tightenlines
\begin{document}
\title{Supersymmetry in the Standard Model}
\author{B. B. Deo}
\address{Physics Department, Utkal University, Bhubaneswar 751 004, India}
\maketitle
       \narrowtext
             \begin{abstract}
We prove that the bosons and massless fermions of one
generation of the standard model are supersymmetric
partners of each other. Except for one additional
auxilliary vector boson, there are no other SUSY particles.

\end{abstract}
\vskip 0.5in

Fayet [1] has constructed a supersymmetric theory of weak, 
electromagnetic and strong interactions. Spin 1/2 gluinoes 
and heavy spin-0 s-quarks are associated with ordinary 
vector gluons and quarks. There is also a class of
s-leptons which include charges ones, a photonic neutrino 
and Higgsinoes. There is a large proliferation of 
elementary particles in SUSY standard models, but there has 
been no experimental signal for any of these SUSY
particles. Yet the standard model [2] without SUSY has been 
remarkably successful in explaining all existing
experimental data of particle physics to a high degree of
accuracy. It has led many physicists to suspect that there
could be a deeper symmetry in the standard model [3]. In
this letter we prove that one such underlying symmetry is
the supersymmetry between the bosons and massless fermions
of one generation.  

One essential requirement for the existence of
supersymmetry is that the fermionic and bosonic degrees of
freedom should be equal. There are the three coloured left
handed doublets $( u_{L}, d_{L})$, two singlets $u_{R}$ and
$d_{R}$ of quarks, one left handed lepton doublet
$(\nu_{L}, e_{L}^{-})$ and the right handed electron
$e_{R}$. Thus fifteen two component fermions in one
generation exist in nature. The bosons are fourteen in all,
the eight gluons $(G_{\mu})$, three $W$-bosons, one $U_{Y}
(1)$ gauge boson $B_{\mu}$ and a complex Higgs doublet
$\phi_{H}$. The remaining two bosonic degrees of freedom
are supplied by choosing an additional massless gauge field
$A_{\mu}$ which serves as the auxillary field for the
model. If another $\nu_{R}$ is found in nature, another
auxilliary $U(1)$ field will be needed and both can be
accomodated as superpartners. Additional $U(1)$ fields do
not affect the predictions of the standard model. For the
present, we will omit $\nu_{R}$. 

The notations involving so many objects are bound to be
complicated and messy. We simplify them as much as
possible.

Let $V_{\mu}^{l}$, $l = 1, 2, \cdots 13$ denote the 
vector fields. We use $ l = 1, 8$ for the gluon fields, $l
= 9$ for the $U_{Y} (1)$ field, $l = 10, 11, 12$ for the
$W$-meson fields and $ l = 13$ for the auxilliary field. We
shall use the temporal gauge [4] where $V_{0}^{l} = 0$. The
electric and magnetic field strength of each $V_{i}^{l}$
are $E_{i}^{l}$ and $B_{i}^{l}$ respectively. Following
Nambu [5], we construct the combination $F_{i}^{l} =
(E_{i}^{l} + iB_{i}^{l} )/ \sqrt{2}$ which satisfy the
nonvanishing equal time commutation relation
\begin{equation}
[ F_{i}^{\dagger l} (x), F_{j}^{m} (y) ] = i \delta_{lm}
\epsilon_{ijk} \partial^{k} \delta (x - y) 
\end{equation}
As suggested by Nambu [5], we construct Wilson line
intergrals to convert the ordinary derivatives acting on
fermionic fields to respective gauge covariant derivatives.

The colour phase function is
\begin{equation} 
U_{C} (x) = \exp ( i g \int_{0}^{x} \sum_{{l} = 1}^{8}
\lambda^{l} V_{i}^{l} dy_{i} ) 
\end{equation}
Denoting 
\begin{equation} 
Y(x) = g^{\prime} \int_{0}^{x} B_{i} dy_{i},
\end{equation}
the isospin space phase functions are
\begin{equation} 
U_{Q} (x) = \exp \left ( \frac{i}{2} g \int_{0}^{x}
\vec{\tau} \cdot \vec{W}_{i} dy_{i} - \frac{i}{6}
Y(x)\right ), 
\end{equation}
\begin{equation} 
U (x) = \exp \left ( \frac{i}{2} g \int_{0}^{x} \vec{\tau}
\cdot \vec{W}_{i} dy_{i} - \frac{i}{2} Y(x)\right ), 
\end{equation}
\begin{equation} 
U_{1}(x) = \exp \left ( - \frac{2i}{3} Y(x) \right ), 
\end{equation}
\begin{equation}
U_{2} (x) = \exp \left ( \frac{i}{3} Y(x) \right ), 
\end{equation}
and
\begin{equation} 
U_{R} (x) = \exp ( i Y(x) ) 
\end{equation}
The $\lambda$'s are Gellmann's SU(3) matrices and the
$\tau$'s are the SU(2) isospin matrices. All these phase
functions are necessary to convert ordinary derivatives on
fermions to covariant ones.

The $\psi^{l}$ will denote the two component fermions.
${l}$ = 1, 2, $\cdots$ 6 refer to the coloured quark
doublet and the sum of the products will mean
\begin{eqnarray} 
\sum_{l = 1}^{6} F_{i}^{\dagger {l}} \psi^{l} = &&
\sum_{{l} = 1}^{3} ( F_{i}^{{l} \star}, F_{i}^{\star {l} +
3} ) \; \; \left ( \begin{array}{c} \psi^{l} \\ 
\psi^{{l} +3} \end{array} \right ) \nonumber \\
&& = \sum_{l = 1}^{3} (F_{i}^{\star l}, F_{i}^{\star l + 3}
) U_{Q} U_{C} \left ( \begin{array}{c} u_{L}^{l} \\
d_{L}^{l} \end{array}\right ) 
\end{eqnarray}
The singlet phased coloured quarks are 
\begin{equation} 
\psi^{l} = U_{1} U_{C} u_{R}^{l} \; \; {\rm for} \; \; l =
7, 8, 9 
\end{equation}
and
\begin{equation} 
\psi^{l} = U_{2} U_{C} d_{R}^{l} \; \; {\rm for} \; \; l =
10, 11, 12 
\end{equation}
Finally
\begin{equation}
\psi^{13} = U_{R} e_{R}
\end{equation}
The Higgs doublet $\phi$ and the lepton doublet $\psi_{L}$
are phased with Wilson line integrals as 
\begin{equation}
\phi = U \phi_{H}
\end{equation}
\begin{equation}
\psi_{L} = U \left ( \begin{array}{c} \nu_{e} \\ e^{-}
\end{array} \right )_{L} 
\end{equation}
Matrix multiplications are implied everywhere. We denote
the four two by two matrices by $\sigma^{\mu}$, $\mu$ = 0,
1, 2, 3. $\sigma^{0} = I$ and $\vec{\sigma}$'s are the
three Pauli spin martices. The supersymmetric charge is now
in a simple form and is given by 

\begin{equation}
Q = \int d^{3} x \left [ \sum_{l = 1}^{13} \left
(\vec{\sigma} \cdot \vec{F}^{+ l} \psi^{l}(x) \right ) +
\left ( \sigma^{\mu} \partial_{\mu} \phi^{\dagger} (x)
\right ) \cdot \psi_{L} (x) \; + {\cal W} \psi^{1} \right ]
\end{equation}
where ${\cal W} = \sqrt{\lambda} (\phi^{\dagger} \phi -
v^{2})$ is the electroweak symmetry breaking term with
$\sigma = \sqrt{2} v = 246$ GeV. $\psi^{1}$ is choosen as
it is one of a left handed doublet and will not contribute
to the commutator of ${\cal W}$ with $\dot{\phi} = \pi$ in
$\psi_{L}$ term. 
The necessary anticommutator between two charges will be
calculated by using the non-vanishing commutators and
anticommutators, 
\begin{equation}
\left [ \pi_{i}^{\dagger} (x), \phi_{j} (y) \right ] = - i
\delta_{ij} \delta (x-y) 
\end{equation}
and
\begin{equation}
\{ \psi^{\dagger l} (x), \psi^{m} (y) \} = \delta_{lm}
\delta (x - y) 
\end{equation}
The relation
\begin{equation} 
\sigma_{i} \sigma_{j} = \delta_{ij} + i \epsilon_{ijk}
\sigma_{k} 
\end{equation}
is frequently used. After lengthy calculation we get the
result 
\begin{equation}
\{ Q_{\alpha}^{\dagger}, Q_{\beta} \} = ( \sigma_{\mu}
P_{\mu} )_{\alpha \beta} 
\end{equation}
$P_{0} = H$ is the Hamiltonian and $\vec{P}$ is the total
momentum. We obtain the correct result for the standard
model, namely [6] 
\begin{eqnarray} 
H && = \int d^{3} X \left [ \: \left ( \sum_{l = 1}^{13} \:
\frac{1}{2} \left ( \vec{E}^{l^{2}} + \vec{B}^{l^{2}} + i
\psi^{l \dagger} (x) \vec{\sigma} \cdot \vec{\nabla}
\psi^{l} (x) \right ) \right ) \right . \nonumber \\ 
&& \left . + \pi^{\dagger} (x) \cdot \pi (x) +
(\vec{\nabla} \phi)^{\dagger} \cdot \vec{\nabla} \phi + i
\psi_{L}^{+} (x) \vec{\sigma} \cdot \vec{\nabla} \psi_{L}
(x) \; + {\cal W}^{2} \right ]
\end{eqnarray}
\begin{eqnarray}
\vec{P} = \int d^{3} X \: \left [ \left ( \: \sum_{l =
1}^{13} (\vec{E}^{l} \times \vec{B}^{l}) - i \dot{\psi}^{l
\dagger} (x) \vec{\nabla} \psi^{l} (x) \right ) \right .
\nonumber \\ 
\left . \dot{\phi}^{\dagger}(x) \cdot \vec{\nabla} \phi (x)
+ ( \vec{\nabla} \dot{\phi}^{\dagger} (x)) \cdot \phi (x) -
i \psi_{L}^{\dagger} (x) \vec{\nabla} \psi_{L} (x) \right ]
\end{eqnarray}
This proves the proposed supersymmetry [7]. The
infinitesmal transformations on the fields can be computed
from the charge. Let $\epsilon$ represent a constant
amticommuting infinitesmal Mojarana spinor. Using the
relation for any field $\varphi$ 
\begin{equation}
\delta \varphi = [ \tilde{\epsilon} Q + Q^{\dagger}
\epsilon, \varphi ]
\end{equation}
we have
\begin{equation}
\delta \psi^{l} = \vec{\sigma} \cdot \vec{F}^{l} \epsilon +
\delta_{l1} {\cal W} \epsilon 
\end{equation} 
\begin{equation} 
\delta F_{i}^{l} = - i \tilde{\epsilon} ( \vec{\sigma}
\times \vec{\nabla} )_{i} \psi^{l} 
\end{equation}
\begin{equation}
\delta \psi_{L} = \sigma^{\mu} \partial_{\mu} \phi \epsilon
\end{equation}
\begin{equation} 
\delta \phi = - i \tilde{\epsilon} \psi_{L} 
\end{equation}
A basic fact about supersymmetry is that the commutator of
two supersymmetry transformations give a spatial
translation [8]. In our case for any field $\varphi$ 
\begin{equation}
[ \delta_{1}, \delta_{2} ] \varphi = \delta_{1} \delta_{2}
\varphi - \delta_{2} \delta_{1} \varphi = a^{\mu}
\partial_{\mu} \varphi 
\end{equation}
where
\begin{equation} 
a^{\mu} = 2 i \tilde{\epsilon}_{1} \sigma^{\mu}
\epsilon_{2} 
\end{equation} 
The equations (27) and (28) are easily obtained by using
Jacobi identity and equations (19) and (22). Thus the
twelve vector bosons of the standard model have the
coloured quarks as their superpartners. The Higgs doublet
is the superpartner of the lepton doublet. The auxillary
vector has the right handed electron as its superpartner.
There are no other additional SUSY particles.
 
We conclude by stating that the massless bosonic and
fermionic fields of the standard model satisfy a closed
supersymmetry algebra. For the first time, such a powerful
deeper symmetry of the physical world of elementary
particles has been discovered and we hope that this will
elucidate the underlying geometry and structure of the
standard model.
 
We have profited from discussions with N. Barik, L. P.
Singh and L. Maharana of the Department of Physics, Utkal
Univeristy. We thank the Director of the Institute of
Physics, Bhubaneswar for providing computer and library
facilities. 
\centerline{REFERENCES}
\begin{enumerate}
\item P. Fayet, Phys. Lett. {\bf 64B}, 159 (1976).
\item S. L. Glashow, Nucl. Phys. {\bf 22}, 579 (1961); A.
Salam and J. C. Ward, Phys. Lett. {\bf 13}, 321 (1965); S.
Weinberg, Phys. Rev. Lett. {\bf 19}, 1264 (1967); A. Salam
in Elementary Particle Theory, Ed. N. Svartholm (Almquist
and Wikskells, Stockholm, 1969) p. 367; S. L. Glashow, J.
Iliopoulos and L. Maiani, Phys. Rev. {\bf D2}, 1285 (1970).
\item One of the earliest to suggest is Veltman in M.
Veltman, Acta Phys. Polon, {\bf B12}, 437 (1981). 
\item N. H. Christ and T. D. Lee, Phys. Rev. {\bf D22}, 939
(1980).
\item Y. Nambu, BCS mechanism, Quasi-supersymmetry and
fermion masses (unpublished); Supersymmetry and Quasi
supersymmetry, Essay in honour of M. Gellmann (1991), EFI
preprints 90-46, 89-08 (unpublished). 
\item See e.g. for the Lagrangian, R. N. Mohapatra in Gauge
theories of fundamental interaction, Edited by R.
N.Mohapatra and C. H. Lai, World Scientific, Singapore
(1981). 
\item See e.g. P. West, Introduction to Supersymmetry and
Supergravity, World Scientific (1986). 
\item J. Wess and B. Zumino, Nucl. Phys. {\bf B70}, 39
(1974). 
\end{enumerate}
\end{document}